# Negative $T$ from a dynamical left-handed neutrino mass

B. Holdom
*Department of Physics*
*University of Toronto*
*Toronto, Ontario*
*Canada, M5S 1A7*
`holdom@utcc.utoronto.ca`

We show how a dynamical Majorana mass for a fourth family left-handed neutrino can make a significant negative contribution to the electroweak correction parameter $T \equiv \Delta\rho/\alpha$ without making a large contribution to $S$ or $U$. We also comment on other possible contributions to $T$ in the context of dynamical symmetry breaking.

The weak interaction parameter $T \equiv \Delta\rho/\alpha$ tends to put strong constraints on new physics, the most notable coming from the fact that the mass splitting within a new weak fermion doublet makes a positive contribution to $T$. Another example is provided by possible new physics hinted at by the current electroweak data. A new massive gauge boson with flavor dependent couplings can mix with the $Z$ and cause shifts in the $Z$ couplings to the various flavors. In particular a new boson coupling to the third family[1] but not the light families can account [1] for the anomalously large values of $\Gamma_b$ and $\alpha_s$ measured at LEP. Variations on this theme have recently been considered for other possible anomalies as well [2]. But the mass mixing between the $Z$ and a heavier boson $X$ is constrained since it will cause the physical $Z$ mass to decrease, resulting in another positive contribution to $T$.

But it should be kept in mind that mixing in the gauge boson kinetic terms is also possible, and this will have the opposite effect on $T$. If the mixing terms appearing in the Lagrangian take the form

$$xm_Z^2 Z'_\mu X'^\mu - 1/2 y Z'_{\mu\nu} X'^{\mu\nu} \tag{1}$$

then upon transforming to mass eigenstates with conventional kinetic terms we find to lowest order in $x$ and $y$ that

$$X'_\mu = X_\mu + r(y-x)Z_\mu \tag{2}$$
$$Z'_\mu = Z_\mu + (rx-y)X_\mu \tag{3}$$

with $r \equiv (m_Z/m_X)^2$. Accounting for effects at second order in $x$ and $y$ as well [3], the contribution to $T$ at lowest order in $r$ is

$$\alpha T = r(x^2 - y^2). \tag{4}$$

In [1] the source of the mixing is a $t$-loop, and the $X$ has an axial coupling to the $t$. In that case $y$ is fairly small compared to $x$, and the result is a positive contribution to $T$ (of order 0.4). In other situations the kinetic mixing could dominate and produce a negative contribution to $T$.

The main topic of this note is the electroweak corrections induced by a Majorana mass for a left-handed neutrino $\nu_L$, a member of a new weak lepton doublet in addition to the three known lepton doublets. As far as we know, all other analyses of electroweak corrections induced by neutrino masses assume either a Dirac neutrino mass or a see-saw pattern in which—in the weak interaction basis—there is a Dirac mass, a Majorana mass for the right-handed neutrino, and no Majorana mass for the left-handed neutrino. $T$ in this framework was calculated in [4], and $S$, $T$, and $U$ were studied in more detail in [5]. $T$ can receive a negative contribution in the see-saw picture, but since the Dirac mass is bounded by the weak scale, the magnitude of the negative contribution to $T$ is quite constrained [4, 5]. A pair of lepton doublets with nonstandard but anomaly-free couplings has also been considered in a see-saw picture [6].

---

[1] Anomalies are canceled since the new gauge boson has equal and opposite couplings to a heavy fourth family.

The restriction to the Dirac or see-saw scenarios is clear in theories with elementary scalar fields. A Majorana $\nu_L$ mass from a vacuum expectation value $v_M$ of a $SU(2)_L$-triplet scalar field would imply a tree level contribution to $\alpha T \approx -(v_M/125 \text{ GeV})^2$. If we are considering a new neutrino, then $v_M$ must be large enough to produce a neutrino mass greater than $m_Z/2$. Unless the Yukawa coupling is very large, the result for $T$ is absurdly negative.

We therefore turn to the alternative—dynamically generated lepton masses. Dynamical masses for fermions beyond the known three families are of interest for dynamical electroweak symmetry breaking. Of special interest for our discussion is the case when the right-handed neutrinos are effectively absent in the TeV theory. In particular, right-handed neutrinos much more massive than a TeV would essentially decouple from the TeV dynamics, and this would leave any left-handed neutrinos beyond the known three to develop dynamical Majorana masses—which must exceed $m_Z/2$. We believe that this picture is a more plausible outcome of strong dynamics than the see-saw picture.

If the extra left-handed neutrinos are technineutrinos in a technicolor theory then they would have to be in a real representation of the technicolor gauge group. Alternatively, the neutrino mass could form in association with the breakdown of a new strong gauge symmetry, which would otherwise have prevented the mass from forming. We note that a similar phenomena has been proposed for the fourth family quark masses, in connection with producing a large $t$ mass [7]. In the following we will simply consider one extra lepton doublet $(\nu_L, E_L)$ and $E_R$ as part of a fourth family, and consider the effects of a dynamical Majorana $\nu_L$ mass on $S$, $T$, and $U$.

If the left-handed Majorana mass is dynamical, then we expect a momentum dependent mass function which falls to zero in the ultraviolet. We shall model this momentum dependent mass by using a constant mass in the presence of an ultraviolet cutoff. We then apply standard Feynman rules for Majorana fermions [8]. In the one-loop diagrams we shall consider, the only change from Dirac mass Feynman rules is in the vertices used. In making the transition from a left-handed projection of a Dirac spinor $\nu_L$ to a self-conjugate four-component Majorana spinor, $N$, the charged current vertex remains the same,

$$\bar{\nu}_L \gamma_\mu E_L \to \bar{N}_L \gamma_\mu E_L, \tag{5}$$

whereas the neutral current is rewritten as

$$\bar{\nu}_L \gamma_\mu \nu_L \to -{}^1\!/_2 \bar{N} \gamma_\mu \gamma_5 N. \tag{6}$$

The factor of 1/2 will be canceled by the two possible contractions involving the vertex (as for a real scalar field), but there remains a symmetry factor of 1/2 for the loop diagram involving the two identical $N$'s.

The contributions to $S$, $T$, and $U$, from the leptons ($\nu$, $E$) may be written in terms of vacuum polarizations of left- and right-handed currents $\Pi_{LL}^{\nu\nu}(q^2)$, $\Pi_{LL}^{EE}(q^2)$, $\Pi_{LL}^{\nu E}(q^2)$ and $\Pi_{LR}^{EE}(q^2)$ [9],

$$\Pi_{LL}^{ab} = -\frac{1}{4\pi^2}\int_0^1 dx \ln\left[\frac{\Lambda^2}{M^2 - x(1-x)q^2}\right][x(1-x)q^2 - \tfrac{1}{2}M^2] \tag{7}$$

$$\Pi_{LR}^{ab} = -\frac{1}{4\pi^2}\int_0^1 dx \ln\left[\frac{\Lambda^2}{M^2 - x(1-x)q^2}\right][\tfrac{1}{2}m_a m_b] \tag{8}$$

$$M^2 = x m_a^2 + (1-x) m_b^2 \tag{9}$$

where $\Lambda$ is the ultraviolet cutoff. In the Dirac mass case the $\Lambda$ dependence cancels in $S, T$, and $U$. For further details on this case we refer the reader to [10], which includes a discussion on the application of the constant mass approximation to dynamical masses and an estimate of possible contributions from pseudo-Goldstone bosons.

To change from the Dirac mass case to the Majorana case ($\nu_L$ to $N$) we make the replacements

$$\Pi_{LL}^{\nu E}(q^2) \to \Pi_{LL}^{NE}(q^2) \tag{10}$$

$$\Pi_{LL}^{\nu\nu}(q^2) \to \tfrac{1}{2}\Pi_{AA}^{NN}(q^2) = \Pi_{LL}^{NN}(q^2) - \Pi_{LR}^{NN}(q^2) \tag{11}$$

to obtain

$$T = \frac{\pi}{s^2 c^2 M_Z^2}\,\mathcal{A}(0) \tag{12}$$

$$S = \frac{4\pi}{M_Z^2}\,[\mathcal{B}(M_Z^2) - \mathcal{B}(0)] \tag{13}$$

$$U = \frac{4\pi}{M_Z^2}\,[\mathcal{A}(M_Z^2) - \mathcal{A}(0)] \tag{14}$$

$$\mathcal{A} = 2\Pi_{LL}^{NE} - \Pi_{LL}^{NN} - \Pi_{LL}^{EE} + \Pi_{LR}^{NN} \tag{15}$$

$$\mathcal{B} = \Pi_{LL}^{NN} - \Pi_{LL}^{EE} - 2\Pi_{LR}^{EE} - \Pi_{LR}^{NN} \tag{16}$$

These expressions are applicable even when the fermion masses are not large compared to $M_Z$.

The extra $\Pi_{LR}^{NN}$ term in $\mathcal{A}$ and $\mathcal{B}$ is due to the Majorana mass. It arises because of a diagram in which one Majorana mass insertion appears on both of the internal fermion lines. This leads to a $\Lambda$ dependence in $T$ but not in $S$ and $U$, and it is clear from (8) that the new contribution to $T$ is negative and proportional to $-\ln(\Lambda/m_N)$. In a theory with elementary scalar fields this $\Lambda$ dependence would be canceled by matching to a renormalizable term in the underlying theory, namely the gauged kinetic term for the $SU(2)_L$-triplet scalar field. In dynamical symmetry breaking there is no such field, and the $\Lambda$ cutoff corresponds to a physical cutoff supplied by the momentum dependence of the neutrino mass function.

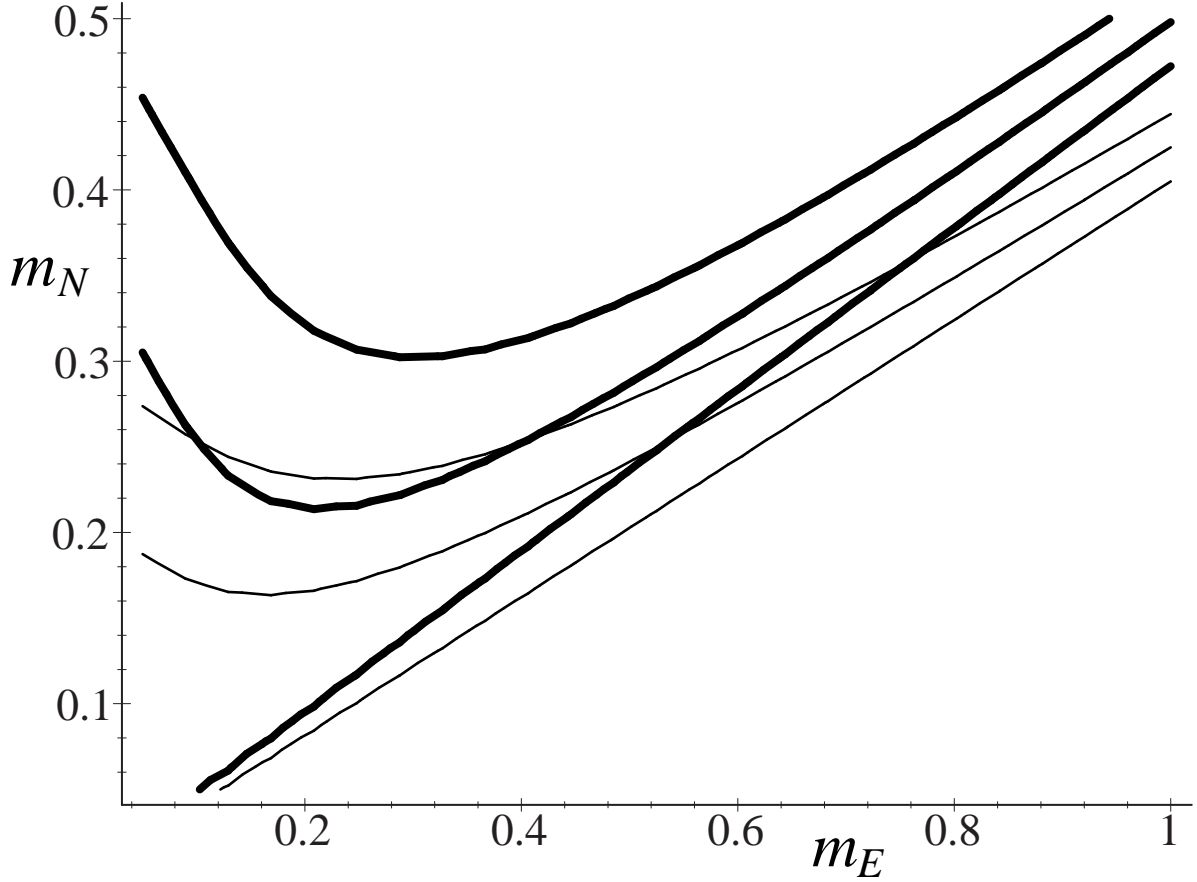

Figure 1. Lines of constant $T$ as a function of the $N$ and $E$ masses in TeV. Thick and thin lines are for $\Lambda = 1.5 m_N$ and $\Lambda = 2 m_N$ respectively. In each case, from top to bottom, $T = -2, -1, 0$.

We therefore estimate $\Lambda$ as follows. Since we have isolated the $\Lambda$ dependence to that appearing in the quantity $\Pi_{LR}(0)$ (which is equivalent to the Goldstone-boson decay constant), we apply the standard Pagels-Stokar approximation [11] to this quantity. This accounts for the momentum dependence of the fermion mass $\Sigma(q^2)$ appearing in the loop, and amounts to fixing the $\Lambda$ dependence appearing in $\Pi^{NN}_{LR}(0)$ from (8) as follows,

$$\ln\left(\frac{\Lambda^2}{m_N^2}\right) \to \int_0^\infty dw\, w\, \frac{S(w)^2 - w[S(w)^2]'/4}{(w + S(w)^2)^2} \qquad (17)$$

with $S(w) \equiv \Sigma(q^2)/m_N$, $w \equiv q^2/m_N^2$, $S(1) \equiv 1$. We find that $S(w) = 2/(1+w)$ corresponds to $\Lambda \approx 1.5 m_N$ while a more slowly falling function such as $S(w) = 10/(9+w)$ corresponds to $\Lambda \approx 2 m_N$. We will consider both of these values for $\Lambda$.

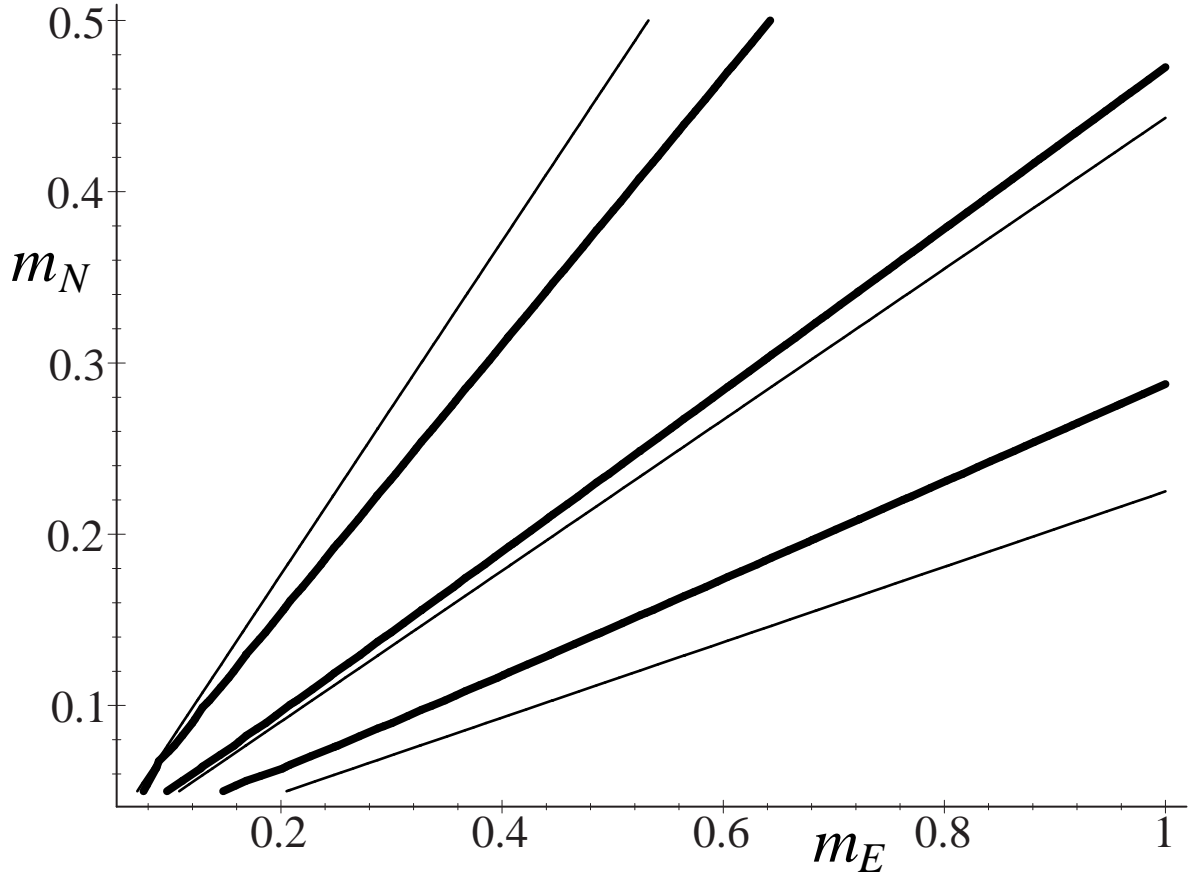

Figure 2. Thick and thin lines are lines of constant $S$ and $U$ respectively as a function of the $N$ and $E$ masses in TeV. From top to bottom in each case $S = 1/6\pi$, 0, $-1/6\pi$ and $U = -1/12\pi$, 0, $1/6\pi$.

We present the contributions to $T$ (Fig. 1) and $S$ and $U$ (Fig. 2) from the lepton doublet as a function of $m_N$ and $m_E$. In Fig. 1 the thick (thin) lines correspond to $\Lambda = 1.5$ ($\Lambda = 2$), and the lines in each case from top to bottom correspond to $T = -2, -1$, and 0. The $T = 0$ line corresponds to where the positive contribution to $T$ from the $N$-$E$ mass splitting cancels the negative contribution from the Majorana $N$ mass. We see that negative values of $T$ occur over a wide range of masses. In Fig. 2 the thick lines show $S = 1/6\pi$, 0, $-1/6\pi$ and the thin lines show $U = -1/12\pi$, 0, $1/6\pi$. ($1/6\pi$ is the usual contribution to $S$ from one degenerate, Dirac-mass, doublet.) The fact that the $S$ contribution is negative for $m_N$ sufficiently small compared to $m_E$ was stressed in [10].

The $S$, $T$, and $U$ contributions may all be small simultaneously. In fact it accidently happens that the $S = 0$ line in Fig. 2 coincides with the $T = 0$, $\Lambda = 1.5$ line in Fig. 1. We see that to have substantial negative $T$ without substantial positive $S$, larger values of $m_E$ are preferred. Then for those values of $m_N$ which give interesting negative values of $T$, both $S$ and $U$ are small. Significant negative $S$ along with negative $T$ would only be possible if our estimate of the effective cutoff $\Lambda$ is an underestimate.

We now briefly discuss other contributions to $T$ in the context of dynamical electroweak symmetry breaking. The problem with $T$ is especially severe in extended technicolor models which introduce new isospin-breaking physics close to a TeV, in order to generate the large $t$ mass. It has become abundantly clear [12] that this new physics produces unacceptably large, positive contributions to $T$; the problem may be expected whenever new TeV mass gauge bosons with isospin-violating couplings have substantial couplings to the TeV mass fermions. Realistic theories must not have such gauge bosons.

Even without such gauge bosons we can identify a minimal contribution to $T$ which arises as long as the $t$-mass is being fed down from a TeV mass weak doublet, say $Q = (U, D)$, via some effective 4-fermion operator. The $b$ mass comes from a correspondingly smaller 4-fermion operator. Let us consider *only* the effects of the $t$-mass operator. (See [7] for a situation in which other isospin-violating 4-fermion operators are suppressed, since they do not have the same anomalous scaling enhancement as the $t$-mass operator.) The point is that the isospin breaking inherent in the $t$-mass operator will feed back and split the $U$ and $D$ masses. Two insertions of the $t$-mass operator (on a $t$-loop) will yield the operator $(1/\Lambda_\Delta^2)\bar{Q}_L Q_R \sigma_{3R} \bar{Q}_R Q_L$, which in turn contributes to $\Delta m$. The result is a positive contribution to $T$:

$$T \propto \frac{\Delta m^2}{f^2} \approx \frac{\langle \bar{Q}Q \rangle^2}{f^2 \Lambda_\Delta^4} \approx \frac{(4\pi)^2 f^4}{\Lambda_\Delta^4} \tag{18}$$

This $\Lambda_\Delta$-operator implies another contribution to $T$ since it induces a small isospin conserving coupling of the $Z$ to $Q_L$. An additional contribution to the $Z$ mass is then generated by a $Q$ loop with two insertions of the new $Z$ coupling.[2] This will have the same sign as the main contribution to the $Z$ mass from the standard $Q$ loop, and the resulting increased $Z$ mass implies a negative contribution to $T$. But the new coupling of the $Z$ to $Q$ is proportional to $f^2/\Lambda_\Delta^2$, and thus the resulting contribution to $T$ will lack the factor of $(4\pi)^2$ appearing in (18). The contribution in (18) will dominate and thus the total contribution to $T$ originating from the $t$-mass operator is expected to be positive.

In summary, theories attempting to explain the $t$ mass in a dynamical context generally produce positive contributions to $T$, in conflict with data. We are suggesting that this predicament may have implications for the origin of neutrino mass.

---

[2] Another way to describe this contribution is to say that two insertions of the $\Lambda_\Delta$-operator generates the operator $(\bar{Q}_R \sigma_3 \gamma_\mu Q_R)^2$, which in turn contributes to the $Z$ mass.

**Acknowledgements**

I thank V. Miransky for discussions. This research was supported in part by the Natural Sciences and Engineering Research Council of Canada.